\documentclass[usenatbib]{mn2e}
\usepackage{psfig}

\title[The effect of starspots on eclipse timings of binary stars]
{The effect of starspots on eclipse timings of binary stars}

\author[C.\,A.\ Watson and V.\,S.\ Dhillon] {C.\,A.\
Watson,\thanks{E-mail: c.watson@sheffield.ac.uk}~and V.\,S.\
Dhillon\\ 
Department of Physics and Astronomy, University of Sheffield,
Sheffield S3 7RH, UK\\}

\date{\center{\Large Accepted for publication in the Monthly 
Notices of the Royal Astronomical Society \\ 
\vspace{.5cm} 25th February 2004}} 

\begin{document}
\maketitle

\begin{abstract}
We investigate the effects that starspots have on the light curves of
eclipsing binaries and in particular how they may affect the accurate
measurement of eclipse timings. Concentrating on systems containing a
low-mass main-sequence star and a white dwarf, we find that if
starspots exhibit the Wilson depression they can alter the times of
primary eclipse ingress and egress by several seconds for typical
binary parameters and starspot depressions. In addition, we find that
the effect on the eclipse ingress/egress times becomes more profound
for lower orbital inclinations. We show how it is possible, in
principle, to determine estimates of both the binary inclination and
depth of the Wilson depression from light curve analysis

The effect of depressed starspots on the O--C diagrams of eclipsing
systems is also investigated. It is found that the presence of
starspots will introduce a `jitter' in the O--C residuals and can
cause spurious orbital period changes to be observed. Despite this, we
show that the period can still be accurately determined even for
heavily spotted systems.

\end{abstract}


\section{Introduction}
\label{sec:wilsonintro}

The study of eclipsing binary stars provides our best source of
accurate stellar parameters, such as masses and radii, which are
essential to theories of stellar structure and evolution. One can also
use eclipsing systems to study, in great detail, the orbital period
evolution of binary stars -- which in turn can be used to test
theories of binary star evolution.

Measurements of the orbital period can be made by timing the eclipse
ingress and egress, and then determining a standard time marker ({\rm
e.g.} the time of mid-eclipse) from which a linear ephemeris can be
calculated. The method of searching for orbital period changes then
amounts to comparing observed time markers with those predicted from
the ephemeris using a standard O--C analysis, and then looking for any
systematic differences. These measurements can be especially accurate
for close binary systems where a compact object such as a white dwarf
is eclipsed, as these provide sharp eclipse transitions. Furthermore,
with the advent of ultra-fast CCD cameras such as {\sc ultracam} (see
\citealt{dhillon01b}; \citealt{dhillon04}), timings of eclipse
ingress/egress events in these objects have been measured to an
accuracy of $\sim$0.1 seconds.

In many of the short-period binaries containing a low-mass
main-sequence star (the secondary star) eclipsing a compact object
(the primary star), the secondary star is either known to show high
levels of magnetic activity ({\rm e.g.} the pre-cataclysmic variable
V471 Tau -- see \citealt{ramseyer95}) or has all the necessary
prerequisites for a magnetically active star (rapid rotation coupled
with a convective envelope). TiO studies by \citet{oneal98} have shown
that the spot coverage on active stars can exceed 50 per cent of the
stellar surface. Combine this with the fact that observations of
sunspots show that they are depressed below the surrounding
photosphere by 100's of kilometres (the so-called `Wilson depression')
then we may well expect the surfaces of such stars to appear heavily
pitted. Therefore, it is plausible that the appearance of a starspot
on the limb of the secondary star as it occults the primary could then
affect the time of eclipse ingress or egress and hence introduce
spurious orbital period changes.

In this paper we show how the presence of starspots and the Wilson
depression can influence eclipse light curves. We start in
Section~\ref{sec:model} with a description of the model that was used
to investigate the effect, and present simulated eclipse light curves
in Section~\ref{sec:results} which demonstrate its magnitude. In
Section~\ref{sec:observe} we investigate the consequences that
depressed starspots may have on O--C analysis.  Finally, in
Section~\ref{sec:conclusions} we discuss the implications that our
results have.

\section{Model}
\label{sec:model}

\begin{figure*}
\centerline{\psfig{figure=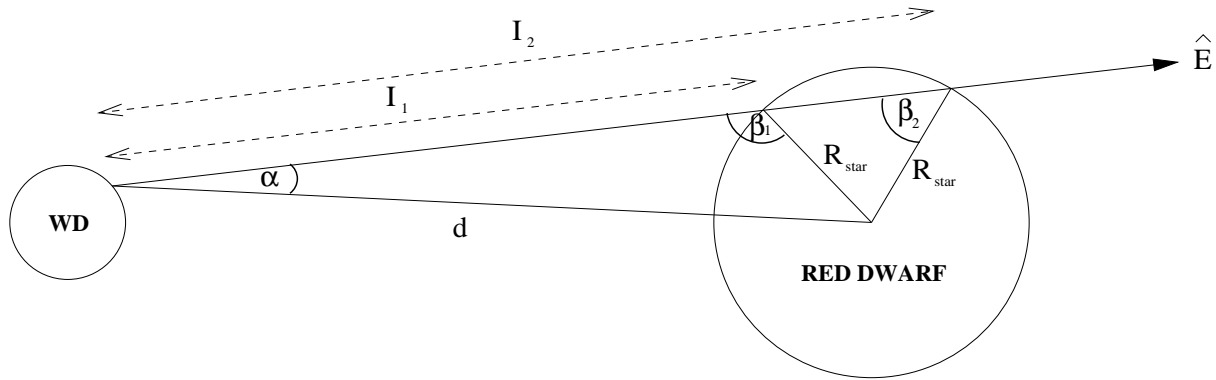,width=16.cm,angle=0.}}
\caption{A schematic (not to scale) showing how the intersection
points with the surface of the red dwarf of the Earth vector
($\hat{E}$) originating from the white--dwarf surface are found. The
white--dwarf is on the left and is marked WD, the red dwarf is on the
right. See text for further details.}
\label{fig:illus}
\end{figure*}

Detached eclipsing binary systems containing one active, late
main--sequence star in orbit around a compact object probably provide
the best candidates in which to detect the effects of
Wilson--depressed starspots on eclipse light curves.  There are three
main reasons for this. First, the sharp eclipse transitions in such
systems allow accurate period measurements to be determined and, in
addition, the compact object acts as a fine pencil--beam able to probe
through a moderately sized, depressed starspot.

Second, detached systems are also desirable as mass transfer between the
components in semi--detached systems makes secure identification of
the ingress and egress times of the compact object more difficult due
to, for example, the presence of an accretion disc. Mass transfer
itself also causes orbital period changes which further complicate the
subsequent period analysis.

Third, the effects that Wilson--depressed starspots have on orbital
period analysis (see later) are insignificant in comparison to the
Applegate effect \citep{applegate92}. This  mechanism requires an
energy $\Delta E \sim (\Delta J)^2/2I$ to transfer angular momentum
$\Delta J$ within the star. $I$ is the moment of inertia of the outer
parts of the star and scales as $MR^2 \sim M^3$ for late--type
stars. The energy available for effecting the change must come from
the star's luminosity which scales as $\sim M^{-4}$ for low--mass
stars. Therefore, the timescale for a given $\Delta J$ scales as $\sim
M^{-7}$ in the Applegate mechanism, and therefore this effect should
be weak in low-mass M dwarf stars \citep[private communication] {marsh04}.

For these reasons, we only consider a binary system containing a
low-mass main-sequence star and a white dwarf in our model. The white
dwarf is modelled as a sphere of radius $R_{wd}$, the surface of which
is divided into a large number (typically 100000's) of elements of
approximately equal area. The main-sequence star is represented as a
sphere of radius $R_{star}$.  Given the component masses and orbital
period, the orbital separation can then be calculated via Kepler's
third law, thereby defining the gross properties of our model binary
system.

We then assume that there is a single star-spot on the main-sequence
star which is circular and depressed below the surrounding immaculate
photosphere by a depth $W_d$. The spot position is specified in terms
of its latitude ($b$), longitude ($l$, where $l$ = 0$\degr$ is defined
as lying on the leading edge of the secondary star) and its angular
size as subtended at the star's centre. In addition, it is assumed
that the transition from starspot to immaculate photosphere is
immediate, that is, the `walls' of the starspot are vertical (probably
a reasonable approximation in the case of large spots) and that the
atmosphere within the Wilson depression is transparent.

Only the eclipse of the white--dwarf itself is considered in these
simulations. To construct a simulated light curve, we first determine
which surface elements on the white dwarf grid are self-obscured by
the white dwarf. This is done by calculating the angle between the
vector pointing towards the Earth (which we shall call the Earth
vector) and the normal vector to the surface element. Taking a set of
Cartesian coordinates ($x,y,z$) which rotate with the binary, with the
origin at the centre-of-mass of the binary, the $z$-axis lying along
the axis of rotation, the $x$-axis lying along the line joining the
centre-of-masses and pointing towards the main-sequence star and the
$y$-axis forming a right-handed set, the Earth vector $\hat{E}$ is
defined as
\[ \hat{E}_x = \cos \theta \sin i \]
\[ \hat{E}_y = - \sin \theta \sin i \]
\[ \hat{E}_z = \cos i, \]
where $i$ is the inclination of the binary and $\theta = 2 \pi
\phi_{orb}$ (where $\phi _{orb}$ is the orbital phase). If the angle
between the Earth vector and normal vector is greater than 90$\degr$
then that surface element is self-obscured.

For each surface element on the white dwarf grid that is not
self-obscured, we then calculate the closest approach to the centre of
the main-sequence star of the line originating from the surface
element and pointing in the direction of the Earth vector. If the
closest approach of this line is greater than the radius of the
main-sequence star ($R_{star}$) then the surface element is not
eclipsed, whereas if the closest approach is less than $R_{star} -
W_d$ then the surface element is eclipsed. A special case presents
itself when the closest approach falls between the previous two
conditions, since it is then possible that the ray from the surface
element may pass unobscured through the Wilson depression.

To determine whether or not this is the case, the two intersection
points of the ray with the surface of the main-sequence star at a
radius $R_{star}$ must be determined. Only if both intersection points
lie within the latitude and longitude of the spot will the ray pass
through the Wilson depression unobscured. If either one of the
intersection points lies outside a spotted region then the ray must
pass through the 'wall' of the spot and is therefore eclipsed.

Fig.~\ref{fig:illus} illustrates how the two intersection points are
ascertained. The length $d$ of the line joining the element on the
white dwarf to the centre of the red dwarf  is easily calculated, as
is the angle $\alpha$ between this line and the Earth vector
$\hat{E}$. The angles $\beta_1$ and $\beta_2$ are then obtained by
applying the sine rule, which gives
\[ \beta_1 = \sin^{-1}\left( \frac{d\sin\alpha}{R_{star}} \right) \]
\[ \beta_2 = \pi - \sin^{-1}\left( \frac{d\sin\alpha}{R_{star}} \right).\]
Using these angles, the distances $I_1$ and $I_2$ to the two
intersection points can then found from the cosine rule, which gives
\[ I_1 = \sqrt{d^2 + R_{star}^2 - 2dR_{star}\cos{\left(\pi-\alpha-\beta_1\right)}} \]
\[ I_2 = \sqrt{d^2 + R_{star}^2 - 2dR_{star}\cos{\left(\pi-\alpha-\beta_2\right)}}. \]
Extrapolation along the Earth vector from the white dwarf surface
element by a distance $I_1$ and $I_2$ give the Cartesian coordinates
of the two intersection points which can then be checked to see if
they lie within the spotted region. The contributions from the visible
surface  elements are then summed together, taking into account their
projected surface areas. Although we could include limb-darkening and
other intensity variations across the white dwarf in this model, for
the purposes of this work we have assumed that the white-dwarf disc is
featureless.  Furthermore, we do not include any contribution from the
main-sequence star, which is effectively treated as a completely dark
occulting body.

In the next section we use this model to simulate light curves that
demonstrate the effect that starspots can have on eclipse timings.

\section{Results}
\label{sec:results}

For our first simulation, we take the pre-CV NN Ser as an example
system and have adopted the binary parameters listed in
Table~\ref{tab:pars}. We define a single spot located at $l$ =
0$\degr$, $b$ = 0$\degr$ and covering 30$\degr$ of the secondary star
(c.f. the polar spot on the primary of VW Cep, which appears to cover
50$\degr$ -- see \citealt{hendry00}).

The depth of the Wilson depression in sunspots has been measured in
many previous studies. For example, \citet{suzuki67} measured depths
of 600--1100 km, \citet{balthasar83} found the depression to be as
large as 2500 km, whereas some authors suggest that not all sunspots
exhibit a Wilson depression and may even show a `reverse Wilson'
effect (\citealt{bagare98}). Given the range in depths measured for
sunspots and the fact that we cannot reliably predict the magnitude of
the Wilson depression in starspots\footnote{Note, however, that work
by \citet{rajaguru02} suggests that stars cooler than the Sun should
exhibit larger Wilson depressions.}, we have simulated eclipse
light curves assuming a range of spot depths spanning 0--1000 km, in
steps of 100 km. We shall call the case where there are no starspots
the `immaculate' case.

\begin{figure}
\centerline{\psfig{figure=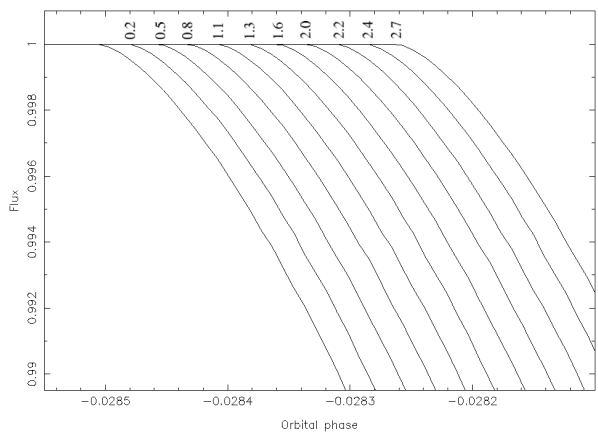,width=7.8cm,angle=0.}}
\caption{A sequence of simulated light curves around the time of white
dwarf ingress. The eclipse assuming an immaculate photosphere is shown
on the left. The other light curves assume that a Wilson--depressed
starspot lies on the occulting limb and the depth of this depression
is increased in 100 km steps from left to right. The time resolution
of the light curves are all 0.1 seconds and the flux has been
normalised by the out-of-eclipse contribution from the white
dwarf. The ingress delay in seconds for each case, as measured against
the immaculate photosphere model, is indicated above the
corresponding light curve ingress point.}
\label{fig:overlay1}
\end{figure}

\begin{figure}
\centerline{\psfig{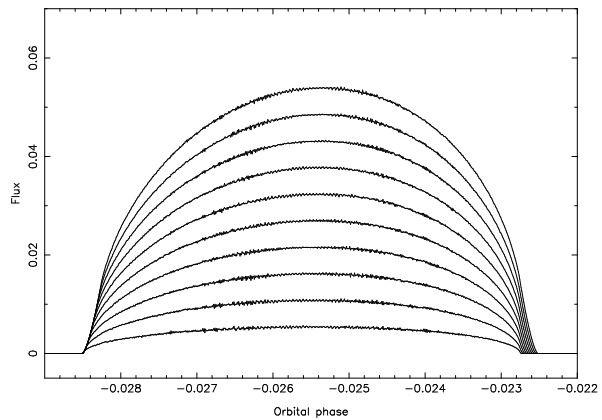}}
\caption{The difference, during ingress, between the immaculate
eclipse light curve and the spotted eclipse light curve. The depth of
the Wilson depression varies from 1000 km (top) to 100 km (bottom) in
steps of 100 km.}
\label{fig:sub1}
\end{figure}

Fig.~\ref{fig:overlay1} shows a sequence of simulated light curves at a
time resolution of 0.1 seconds (as typically obtained by {\sc
ultracam}) around the time of white dwarf ingress.  It can be clearly
seen that the presence of a depressed starspot causes the time of
ingress to be delayed compared to the time of ingress for the
immaculate case. For a starspot with a 1000 km Wilson depression, this
delay amounts to 2.7 seconds in our model. Fig~\ref{fig:sub1} shows
the difference between the immaculate ingress light curve and the
spotted ingress light curve for spots of different Wilson
depressions. In this case we see that the maximum difference between
the immaculate and spotted light curves occurs approximately 37 seconds
($\sim$0.0033 in phase units) after the start of ingress, and for a
Wilson depression of 1000 km this difference amounts to 5.4\% of the
total contribution from the white dwarf.

For our next series of simulations we have varied the orbital
inclination of the system. For each inclination, the spot latitude was
changed such that the centre of the starspot lay approximately on the
position of the limb where the white dwarf was eclipsed. Assuming that
the white dwarf is eclipsed by the leading edge of the secondary star
(i.e., at the position $l$ = 0$\degr$), then the latitude ($b$) of the
region of the secondary that occults the centre of the white dwarf for
any orbital inclination is given by
\[b = \sin^{-1}\left[ \frac{a.\tan\left(90-i\right)}{R_{star}}\right], \]
where $a$ is the binary separation. For each inclination, the latitude
of the spot was varied according to the above equation.

\begin{figure}
\centerline{\psfig{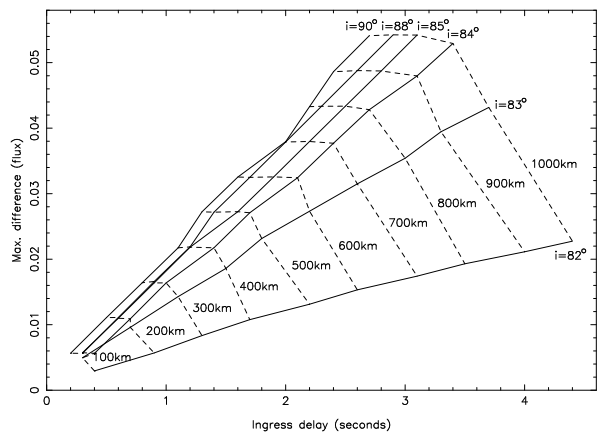}}
\caption{A summary of how starspots of different Wilson depressions
affect both the ingress delay (horizontal axis) and the maximum
difference in flux between the ingress light curve of a spotted and
immaculate star (vertical axis). The solid lines show the effects at
different inclinations and the dashed lines indicate the depth of the
Wilson depression assumed in each model.}
\label{fig:all}
\end{figure}

The results are summarised in Fig.~\ref{fig:all}. This shows that, as
the inclination is lowered, there is a lengthening in the delay of the
eclipse ingress timing for any particular depth of the Wilson
depression. At an inclination of 82$\degr$, this delay can be as much
as 4.4 seconds in the presence of an appropriately placed spot with a
1000 km Wilson depression.

\begin{figure}
\centerline{\psfig{figure=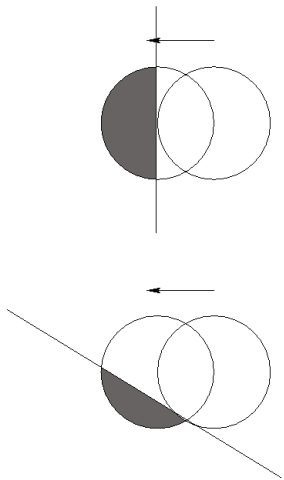,width=5.0cm,angle=0.}}
\caption{A schematic showing how the projected area of the white dwarf
obscured per unit time is smaller for eclipsing systems of lower
inclinations. The vertical line represent the limb of the occulting
star and the circles represent the white dwarf. The top diagram
depicts the white dwarf as it first contacts the companion (right-hand
circle) in a high inclination binary. The left-hand circle depicts the
white dwarf a short time later when the shaded portion has been
eclipsed.  The bottom panel shows the same scenario but for a lower
inclination system. The arrow indicates the motion of the white--dwarf
relative to the occulting limb of the red dwarf which has been
depicted as stationary for clarity.}
\label{fig:schematic}
\end{figure}

Another feature of note is that the maximum difference between the
immaculate and spotted eclipse light curves shows a reverse trend with
inclination compared to the ingress delay times. For lower
inclinations the magnitude of this difference diminishes, which is
due to the changing slope of the occulting limb of the secondary
star. Fig.~\ref{fig:schematic} is a schematic showing how the limb
eclipses the white dwarf at an inclination of 90$\degr$ and also at a
lower inclination. From this it can be seen that the slope of the
ingress light curve should be much shallower in the low inclination
case. Therefore, even though there is a greater ingress delay at a low
inclination, the white dwarf is obscured far more slowly and this
causes the maximum difference between the immaculate and spotted
eclipse light curves to diminish.

This feature (the anti-correlation between the ingress delay and
eclipse difference) is potentially useful. If one can observe an
immaculate eclipse light curve and establish an accurate period for the
system then, by observing any systematic deviations in ingress/egress
timings and simultaneously assessing the maximum difference between
the immaculate and spotted eclipses, it should be possible to uniquely
define a point on a diagram similar to Fig.~\ref{fig:all} that gives
the orbital inclination and size of the Wilson depression.

\begin{table}
\begin{center}
\caption[]{Binary properties assumed in the model, taken from
\protect\citet{catalan94}.}
\begin{tabular}{cc}
\hline Parameter & Value \\ \hline $M_1$/M$_{\odot}$ & 0.57 \\
$M_2$/M$_{\odot}$ & 0.12 \\ $R_1$/R$_{\odot}$ & 0.017 \\
$R_2$/R$_{\odot}$ & 0.153 \\ $P_{orb}$ (s) & 11239 \\ $i\degr$ & 90 \\
\hline \\
\label{tab:pars}
\end{tabular}
\end{center}
\end{table}

\section{Observational practices}
\label{sec:observe}

Given that starspots can alter the times of eclipse ingress and
egress, it is important to understand the effect that this has on
period determination and O--C analysis.

In order to investigate this we have used the model described in
Sections~\ref{sec:model} \&~\ref{sec:results} to simulate a series
of eclipse light curves over many cycles. We assume a series of 12 hour
long observations, each separated by a 30 day gap and all having a
time resolution of 0.1 seconds. Over the 30 day interval between each
12 hour observation it is assumed that the spots evolve such that they
are located in completely different positions. The spot distribution
is modelled as a series of randomly placed spots of 30$\degr$ diameter
covering a total of 30$\%$ of the stellar surface. Each spot is then
randomly assigned a Wilson depression of between 100--1000 kilometres,
leading to a systematic shift of the ingress or egress transition of
0.0--2.7 seconds. During the course of each 12 hour observation both
the spot locations and depths are fixed. We adopt the binary
parameters listed in Table~\ref{tab:pars} throughout the simulations.

\begin{figure}
\centerline{\psfig{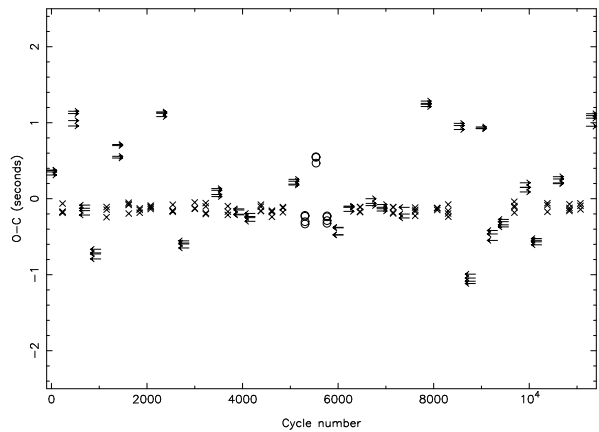}}
\caption{The O--C diagram derived from the simulation described in
Section~\ref{sec:observe}. The left--pointing arrows denote
observations where a spot lies on the trailing limb, right--pointing
arrows denote observations when a spot lies on the leading limb, open
circles where spots lie on both limbs and crosses where both limbs are
clear of spots. Note that 3--4 eclipses are covered in each 12 hour
observation, which leads to the grouping of several points around the
same cycle number, the scatter in each group being due to timing
errors assumed in the simulation.}
\label{fig:ocplot}
\end{figure}

To investigate the effects of starspots, we first determined the time
of mid--eclipse for each light curve by locating the start time of
ingress and the end time of egress for each eclipse in turn. The
mid--eclipse was then defined as lying exactly half--way between these
points. We then fit these times of mid--eclipse over the whole set of
simulated observations by a linear ephemeris of the form
\[HJD = T_0 + nP, \]
where $T_{0}$ is the mid--eclipse zero--point, $n$ is the cycle number
and $P$ is the orbital period. Fig.~\ref{fig:ocplot} shows the O--C
diagram derived from this simulation.  This shows that the presence of
starspots causes a random `jitter' in the O--C diagram. From
Fig.~\ref{fig:ocplot} it is possible to distinguish between starspots
that lie on the leading or trailing limb since a positive O--C
residual results when a spot is on the leading limb, whereas a spot on
the trailing limb results in a negative O--C residual. Where starspots
lie on both limbs the O--C residual is dependent on which spot has the
largest Wilson depression and hence it is more difficult to tell from
Fig.~\ref{fig:ocplot} when this latter situation has occurred.

\begin{figure}
\centerline{\psfig{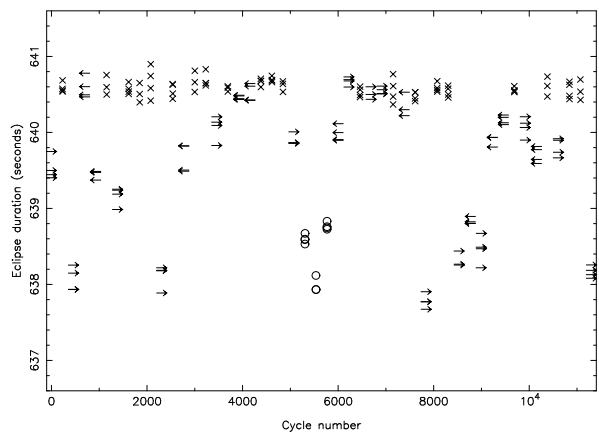}}
\caption{The duration of each eclipse in seconds; the symbols are the
same as in Fig~\ref{fig:ocplot}.}
\label{fig:ecldur}
\end{figure}

\begin{figure}
\centerline{\psfig{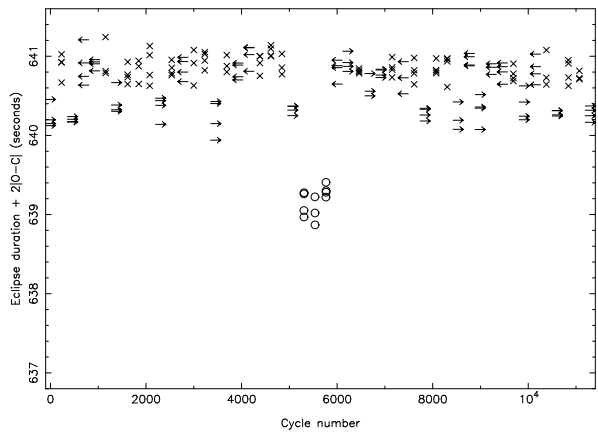}}
\caption{A plot of the eclipse duration $D$ summed with twice the
magnitude of the O--C residuals for that eclipse observation; The
symbols are the same as in Fig~\ref{fig:ocplot}.}
\label{fig:solution}
\end{figure}

One might expect a correlation, however, between the eclipse duration
(measured from the start of ingress to the end of egress and plotted
in Fig.~\ref{fig:ecldur}) and the observed O--C residuals.
This is because the presence of a starspot on the leading limb of the
occulting star will cause a delay in the ingress transition of, for
example, $\delta t$ seconds. This would then cause the observed
eclipse duration $D$ to shorten by $\delta t$ seconds but result in a
smaller shift in the mid--eclipse timing of $\delta t/2$ seconds.
If the precise orbital period and zero--point were known then the
shift in the mid--eclipse timing would also result in an O--C
residual of magnitude $\delta t/2$. Therefore, by plotting
\[D + 2\times|O-C|, \]
points taken from observations where a spot lies on just one limb of
the occulting star should fall on the same line, this is shown in
Fig.~\ref{fig:solution}. In practice, we do not know the precise
period and zero--point, leading to the larger scatter seen in
Fig.~\ref{fig:solution}. Points where starspots are present on both
limbs, however, lie systematically lower since their O--C residuals
are partially cancelled out. Using this method, it is therefore
also possible to identify those observations where starspots lie
on both limbs.

\begin{figure*}
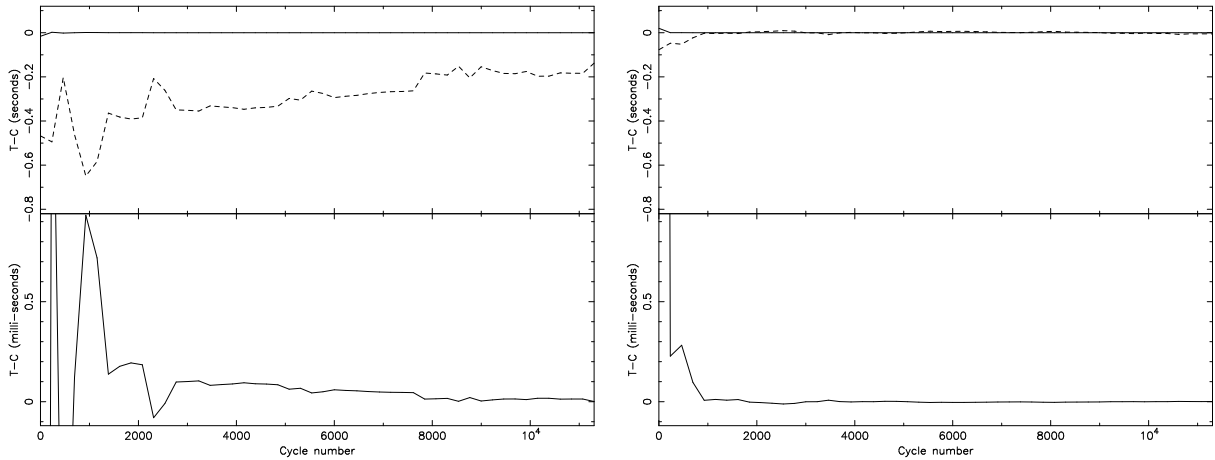

\begin{center}
\begin{tabular}{ll}
\psfig{figure=ocdiff.ps,width=7.8cm,angle=-90.} &
\psfig{figure=ocdiff2.ps,width=7.8cm,angle=-90.} \\
\vspace{0.2cm}\\
\end{tabular}
\end{center}
\caption{The discrepancy between the true orbital period and
zero--point and those computed using a linear ephemeris fit. Top
panel -- the full line indicates the error (true - computed) on the
orbital period, and the dashed lines indicates the error on the
zero-point determination, both in seconds. Bottom panel -- the error
on the orbital period determination in milli-seconds. The left-hand
panels show the results for the spotted model described in Section
4. and the right-hand panels show the results for an unspotted but
otherwise identical model for comparison. It is clear that the orbital
period can still be accurately determined within $\sim$ 3000
cycles despite the influence of starspots. The error in the
zero--point is found to be greater, but is still determined to within
$\sim$0.13 seconds over the full dataset.}
\label{fig:ocdiff}
\end{figure*}

Since we have shown how depressed starspots can affect O--C diagrams,
we now investigate the impact that this has on the accuracy of both
the period and the mid--eclipse zero-point ($T_0$) determined from the
linear ephemeris fit. To do this we fit a linear ephemeris to the
accumulated data after each observation date -- essentially updating
the ephemeris after each observation. We then compared the calculated
period and mid--eclipse zero--point to the true values used in the
simulation in order to asses how accurate the calculated values are,
and how they vary with increasing cycle number. This information is
shown in the left--hand panel of Fig.~\ref{fig:ocdiff}. The
right--hand panel shows a similar analysis but for an immaculate star
with no starspots as a comparison.

As can be seen, the orbital period is quickly determined to better
than 1-ms accuracy after four separate observations (after cycle 695),
even though some ingress and egress times are affected by the presence
of starspots during this time. The reason why spots do not appear to
have an adverse effect on the orbital period determination is that
they only introduce a scatter in the O--C diagram
(Fig.~\ref{fig:ocplot}) and do not change its overall slope -- allowing
the orbital period to still be accurately determined.

The mid-eclipse zero-point ($T_0$) can also be determined with
reasonable accuracy, although it still shows a departure from the true
zero-point of $\sim$0.13 seconds after $\sim$11000 cycles.  Another
feature to note is that the trend in the mid-eclipse zero--point error
appears to reflect that of the error in the orbital period. This is
because a calculated orbital period that is longer than the true
period causes the calculated times of mid--eclipse to become stretched
out. This leads to the determination of an earlier zero--point in
order to compensate for this and minimise the O--C residuals. It is
this that causes the non--spotted O--C residuals in
Fig.~\ref{fig:ocplot} to lie systematically below zero.

It should be noted that, although the overall slope of the O--C
residuals is not changed by the presence of spots, during the
appearance and/or disappearance of a starspot the slope of the O--C
curve is temporarily changed. Therefore, spurious orbital period
changes may be observed, especially over short time baselines. Indeed,
a quadratic ephemeris fit did reveal a spurious but statistically
significant orbital period change. Fit over the entire 1500 day
dataset, we find that the rate of orbital change $\dot{P}$ =
$9.5\times10^{-9} \pm 1.0\times10^{-9}$. In the model simulations
presented here, this amounts to the false detection of a 1.2 second
change in the orbital period over a four year interval. As a further
check, we also produced a simulated dataset using a similar model to
before but assuming zero starspot coverage.  As expected, a quadratic
ephemeris fit to this data did not find a significant period change.

Given that the appearance and disappearance of starspots on the limb
is a largely random event (although there is evidence that starspots
can form at preferred longitudes, see e.g. \citealt{holzwarth03}), the
sense and magnitude of the resultant $\dot{P}$ is also random. It
would, therefore, be distinguishable from other mechanisms such as the
Applegate effect \citep{applegate92} and presence of a third body
which both produce cyclical variations (e.g. \citealt{soydugan03}),
and magnetic braking which will give a steadily decreasing orbital
period. The orbital period change caused by the latter effect would
also be reinforced by the parabolic dependence with time exhibited by
a true orbital period change. In conclusion, the orbital period can
still be accurately determined if the observations are carried out
over a long enough interval over which the random effects of starspots
will cancel out.

\section{Conclusions}
\label{sec:conclusions}

In Section~\ref{sec:results} we showed that depressed starspots can
delay ingress or advance egress in eclipsing main-sequence white-dwarf
binaries by several seconds, and that the magnitude of this effect
becomes stronger for lower inclinations. We have also shown that it
is, in principle at least, possible to determine both the inclination
of the system and the depth of the Wilson depression from light curve
analysis.

Depressed starspots also cause a `jitter' in the residuals of O--C
diagrams, which can also result in the false detection of spurious
orbital period changes. Such changes due to starspots would be
distinguishable from other mechanisms that cause period changes, and
it should still be possible to determine the orbital period accurately.

Observational evidence for such effects are of interest. First, it
would confirm that starspots also show the Wilson depression exhibited
by sunspots -- providing further tests of the solar-stellar
connection.  Second, it is not entirely clear whether the giant
starspots that are found in Doppler images of rapidly rotating stars
are actually monolithic, or groups of far smaller, individually
unresolved, starspots. The effects described here will only be
produced by relatively large starspots, and observations of these
effects provide one of the few ways in which it would be possible to
determine the monolithic nature (or otherwise) of starspots.

The best candidates in which to see the effects of Wilson--depressed
starspots are eclipsing detached white--dwarf/late M--dwarf binaries,
where the Applegate mechanism will be weak. However, even where the
Applegate mechanism is strong, Wilson--depressed starspots would still
cause variations in the eclipse width to be seen, which would otherwise
remain constant under simple period changes.

Finally, \cite{kalimeris02} have shown that intensity variations due
to starspots (not taking into account Wilson depressions) can
introduce disturbances of up to $\sim$0.01 days in the O--C residuals
of contact binaries. Given the rapid evolutionary timescales of spots
(of the order of days) seen in recent Doppler images of the contact
binary AE Phe (\citealt{barnes04}), this may lead to explaining some
of the observed jitter in the O--C curves of these objects. Since, in
this work, the effects of a Wilson depression seem to result in a
scatter of only a few seconds in the O--C residuals, it seems unlikely
that the Wilson depression will be a significant source of jitter for
contact binaries. However, given the different geometries involved and
procedures for determining standard time-markers, a proper assessment
of whether this latter statement holds true is beyond the scope of
this paper.

\section*{\sc Acknowledgements}

CAW is employed on PPARC grant PPA/G/S/2000/00598. The authors
acknowledge the use of the computational facilities at Sheffield
provided by the Starlink Project, which is run by CCLRC on behalf of
PPARC. We are also indebted to Tom Marsh for pointing out, in such
detail, that the Applegate effect should be greatly diminished for
low--mass stars and for casting a critical eye over an earlier draft
of this paper. Many thanks also to the referee, Ron Hilditch, for
his helpful comments.

\bibliographystyle{mn2e}
\bibliography{abbrev,refs}

\end{document}